\documentclass{PoS}
\title{Reconstructing Dark Matter Properties via Gamma-Rays with {\em Fermi}-LAT}

\ShortTitle{Reconstructing Dark Matter Properties via Gamma-Rays with {\em Fermi}-LAT}

\author{\speaker{Nicol\'as BERNAL}
        \thanks{The author was supported by the EU project MRTN-CT-2006-035505 HEPTools and
        the DFG TRR33 `The Dark Universe'.}\\
        Bethe Center for Theoretical Physics and Physikalisches Institut, Universit\"at Bonn\\
        Nussallee 12, D-53115 Bonn, Germany\\
        Centro de F\'isica Te\'orica de Part\'iculas, Instituto Superior T\'ecnico\\
        Avenida Rovisco Pais, 1049-001 Lisboa,
        Portugal\\
        E-mail: \email{nicolas@th.physik.uni-bonn.de}}

\abstract{We study the capabilities of the {\it Fermi}--LAT instrument
for identifying particle Dark Matter properties as mass, annihilation cross section and annihilation channels, with
gamma-ray observations from the Galactic Center. For the potential Dark Matter signal, besides the
prompt gamma-ray flux produced in Dark Matter annihilations, we also take into account the flux
produced by inverse Compton scattering of the electrons and positrons generated in Dark Matter
annihilations off the interstellar photon background. We show that the addition of this contribution 
is crucial in the case of annihilations into $e^+e^-$ and $\mu^+\mu^-$ pairs.
In addition to the diffuse galactic and extragalactic background,
we also consider the full catalog of high-energy gamma-ray point sources detected by {\it Fermi}.
The impact of the degeneracies between the different Dark Matter annihilation channels has been
studied. We find that for Dark Matter masses below $\sim 200$ GeV and for typical thermal annihilation cross sections,
it will be possible to obtain stringent bounds on the
Dark Matter properties.}

\FullConference{Identification of Dark Matter 2010\\
                 July 26 - 30 2010\\
                 University of Montpellier 2, Montpellier, France}

\begin{document}

\section{Introduction}
If Dark Matter (DM) is detected and identified, the measurement of its properties
like mass, annihilation cross-section and annihilation channels plays a central role in
the determination of the particle nature of the DM. It will allow us to constrain
models of particle physics beyond the Standard Model, for instance supersymmetry and universal
extra dimensions. Furthermore a convincing DM discovery may require consistent signals
in multiple experiments in multiple channels (direct, indirect, collider). We
discuss the capabilities of the {\it Fermi}--LAT instrument for identifying particle DM
properties with gamma-ray observations from the Galactic Center (GC).

The differential intensity of the photon signal
from a given observational region
in the galactic halo from the annihilation of DM
particles has different possible origins: internal
bremsstrahlung and secondary photons (prompt) as well as Inverse Compton
Scattering (ICS). External bremsstrahlung and synchrotron emission also
contribute to the photon flux; however, for the energies of interest here
and for typical DM masses, both bremsstrahlung and synchrotron emission
are expected to be subdominant with respect to
ICS. For the sake of simplicity we will
neglect these sources in what follows.

The differential flux of prompt gamma-rays from DM annihilations
and coming from a direction within a solid angle $\Delta\Omega$ is given by
\begin{equation}
\left(\frac{d\Phi_{\gamma}}{dE_\gamma}\right)_{{\rm prompt}} (E_{\gamma},\,
\Delta\Omega) = \frac{\langle\sigma
  v\rangle}{2\,m_\chi^2}\sum_i\frac{dN_{\gamma}^i}{dE_{\gamma}}\, 
\textrm{BR}_i \, \frac{1}{4\,\pi} \, \int_{\Delta\Omega}d\Omega \,
\int_\textrm{los}\rho\big(r(s,\,\Omega)\big)^2 \, ds\,, 
\label{Eq:promptflux}
\end{equation}
where $\langle\sigma v\rangle$ is the total thermally averaged annihilation cross section,
$m_\chi$ the mass of the DM particle, $\textrm{BR}_i$ the annihilation fraction into channel $i$,
$dN_\gamma^i/dE_\gamma$ the differential gamma-ray yield of standard model particles into photons of energy
$E_\gamma$, $\rho(r)$ the DM density profile and $r$ the distance from the GC.
Here we will focus on the NFW halo profile \cite{Navarro:1995iw}; the dependence on
the DM halo profile has been studied in reference \cite{Bernal:2010ip}.

An abundant population of energetic electrons and positrons produced
in DM annihilations either directly or indirectly from the hadronization, fragmentation, and
subsequent decay of the SM particles in the final states, gives rise to
secondary photons at various wavelengths via ICS off the diffuse
radiation fields in the galaxy.
We approximate this photon background as a superposition of three
black-body spectra consisting of the CMB,
the optical starlight and the infrared
radiation due to rescattering of starlight by dust \cite{ics}.
The differential flux of high energy
photons produced by the ICS processes is given by \cite{Blumenthal:1970gc}
\begin{equation}
\left(\frac{d\Phi_{\gamma}}{dE_\gamma}\right)_{{\rm ICS}} (E_{\gamma},\,
\Delta\Omega) = \frac{1}{E_{\gamma}} \, \frac{1}{4\pi} \, 
\int_{\Delta\Omega}d\Omega \, \int_\textrm{los}ds \,
\int_{m_e}^{m_\chi}dE\,\mathcal{P}(E_{\gamma},\,E)
\, \frac{dn_e}{dE}\big(E,\,r,\,z\big) ~,
\label{Eq:ICSflux}
\end{equation}
where $\mathcal{P}(E_{\gamma},\,E)$ is the differential power emitted into scattered photons of energy
$E_{\gamma}$ by an electron with energy $E$.
The minimal and maximal energies
of the electrons are determined by the electron mass $m_e$ and the
DM particle mass.
The quantity $dn_e/dE$ is the electron plus
positron spectrum after propagation in the Galaxy, which will
differ from the energy spectrum produced at the source.  We determine
the propagated spectrum by solving the diffusion-loss equation that describes
the evolution of the energy distribution for electrons and positrons
assuming steady state~\cite{propaga}. Regarding the propagation parameters
(like diffusion coefficient, energy losses and thickness of the diffusion zone),
we take their values from the commonly used MED model~\cite{propaga}.
Again, the dependence on the propagation model has been studied in reference \cite{Bernal:2010ip}.

There are three main components contributing to the high-energy gamma-ray
background: the diffuse galactic emission has been estimated by taking
the conventional model of the GALPROP code~\cite{galprop}.
On the other hand, another source of background particularly important
when looking at the GC is that of resolved point sources.
We consider all the point sources detected by the first
11~months of {\it Fermi}--LAT~\cite{pointsources} lying in the region of interest.
Finally, for the isotropic extragalactic gamma-ray background we used the
recent measurements by the {\it Fermi}--LAT collaboration~\cite{Abdo:2010nz}.
For a $10^\circ \times 10^\circ$ region around the GC, the diffuse galactic emission
dominates below $\sim 20$~GeV. Above that value, the emission coming from point
sources is the most important. The isotropic extragalactic gamma-ray background
is at the percent level.

The Large
Area Telescope ({\it Fermi}--LAT) is the primary instrument on board of
the {\it Fermi Gamma-ray Space Telescope}. It performs an all-sky survey, covering a
large energy range for gamma-rays, with an effective area $\simeq 8000$~cm$^2$
and a field of view of $2.4$~sr.  In the following
analysis, we consider a 5-year mission run, and an energy range from
1~GeV extending up to 300~GeV.  We divide this energy interval into 20
evenly spaced logarithmic bins. 
In order to maximize the signal-to-noise ratio, it has been pointed out that for a NFW
profile the best strategy is to focus on a region around the GC of
$\sim10^\circ\times 10^\circ$~\cite{maxsb}.  Hence, this is our choice.

\section{Reconstructing Dark Matter properties}
Once gamma-rays are identified as having been
produced in DM annihilations, the next step concerns the possibilities
of constraining DM properties \cite{papas}.
\begin{figure}[ht!]
\begin{center}
\includegraphics[width=7.5cm]{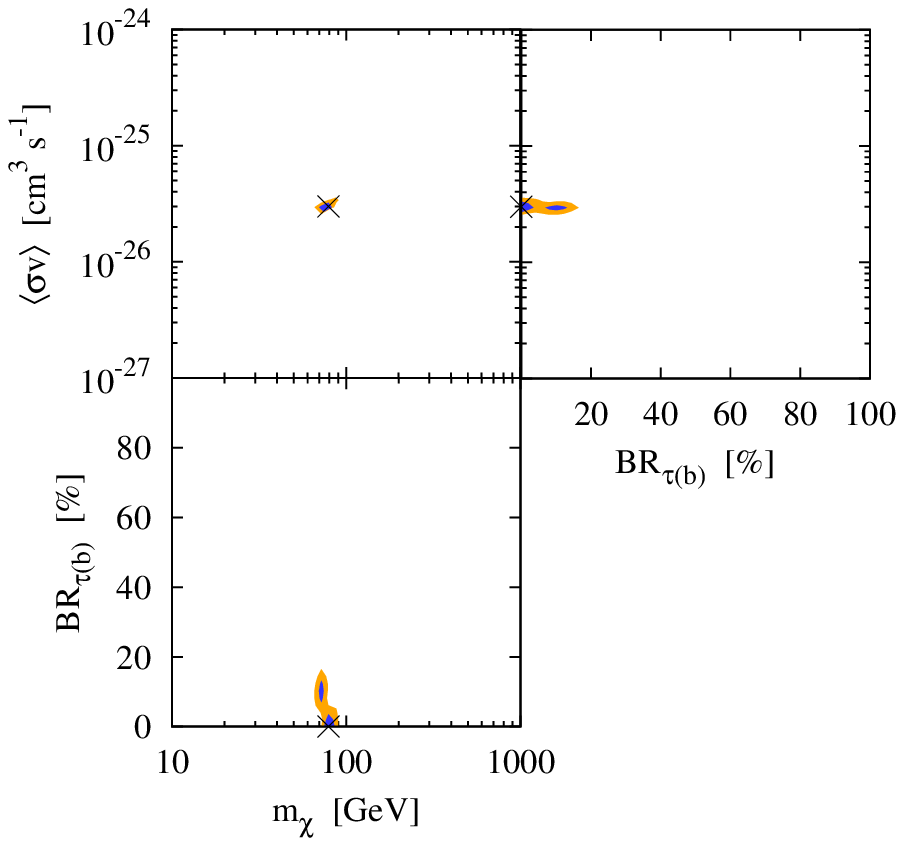}
\includegraphics[width=7.5cm]{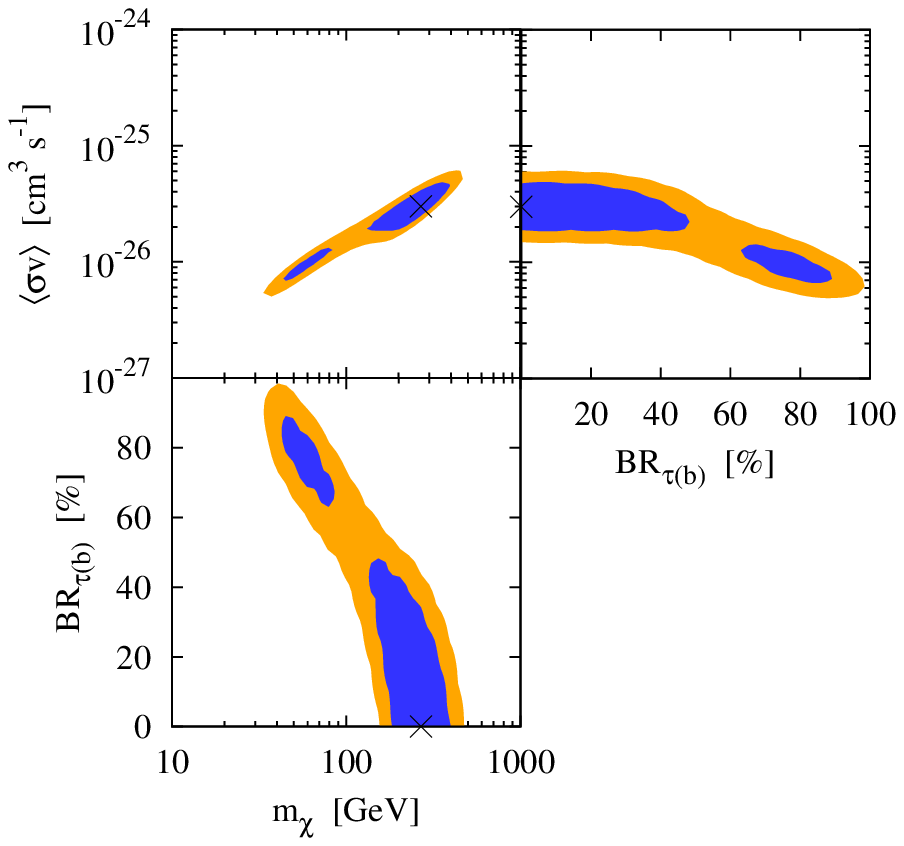}
\end{center}
\caption{{\it Fermi}--LAT abilities to constrain DM properties.  We
  consider DM annihilation into a pure $b\bar b$ final state and
  two DM masses: $m_\chi=80$~GeV (left panels) and
  $m_\chi=270$~GeV (right panels).  Dark blue (light orange) regions
  represent $68\%$~CL ($90\%$~CL) contours.  We assume a
  $10^\circ \times 10^\circ$ observational region around the GC, a NFW DM
  halo profile, the MED propagation model and $\langle\sigma
  v\rangle=3\cdot 10^{-26}$~cm$^3$~s$^{-1}$.  The
  black crosses indicate the values of the parameters for the
  simulated observed ``data''.}
\label{Fig:b10deg}
\end{figure}
In figure~\ref{Fig:b10deg} we depict the {\it Fermi}--LAT reconstruction
prospects after 5~years for DM annihilation into a pure
$b\bar b$ final state reconstructed as either $\tau^+\tau^-$ or
$b\bar b$ and two possible DM masses: $m_\chi=80$~GeV (left panels)
and $m_\chi=270$~GeV (right panels).  We also assume DM
particle with a typical thermal annihilation cross section $\langle\sigma
v\rangle=3\cdot 10^{-26}$~cm$^3$~s$^{-1}$, the MED propagation model,
a NFW DM halo profile and a $10^\circ \times 10^\circ$ observational
region around the GC. These benchmark points are represented in the
figure by black crosses. The dark blue regions and the light orange
regions correspond to the $68\%$~CL and $90\%$~CL contours
respectively.  The different panels show
the results for the planes $(m_\chi,\,\langle\sigma v\rangle)$,
$($BR$_{\tau(b)},\,\langle\sigma v\rangle)$ and
$(m_\chi,\,$BR$_{\tau(b)})$, marginalizing with respect to the other
parameter in each case. BR$_{\tau(b)}=100\%$ ($0\%$) corresponds to an annihilation
into a pure $\tau^+\tau^-$ $\left(b\bar b\right)$ final state.
For the first model chosen in
figure~\ref{Fig:b10deg} (left panels), $m_\chi=80$ GeV and in general for light DM masses,
the reconstruction prospects seem to be promising, allowing the
determination of the mass, the annihilation cross section and the
annihilation channel at the level of $\sim$20\% or better.
On the other hand, for heavier DM particles, the
regions allowed by data grow considerably worsening the abilities of the
experiment to reconstruct DM properties.  This is shown for the second
model in figure~\ref{Fig:b10deg} (right panels), $m_\chi=270$~GeV.  In
this case, {\it Fermi}--LAT would only be able to 
constrain the DM mass to be in the range $\sim (30-500)$~GeV and
determine the annihilation cross section within an order of magnitude.
Let us note the appearance of a second spurious minima corresponding to a 
lighter mass $m_\chi\sim 60$ GeV, an annihilation cross section
$\langle\sigma v\rangle\sim 10^{-26}$ cm$^3$ s$^{-1}$ and annihilating
mainly ($\sim 75\%$) on $\tau^+\tau^-$ pairs.

\begin{figure}[ht!]
\begin{center}
\includegraphics[width=7.5cm]{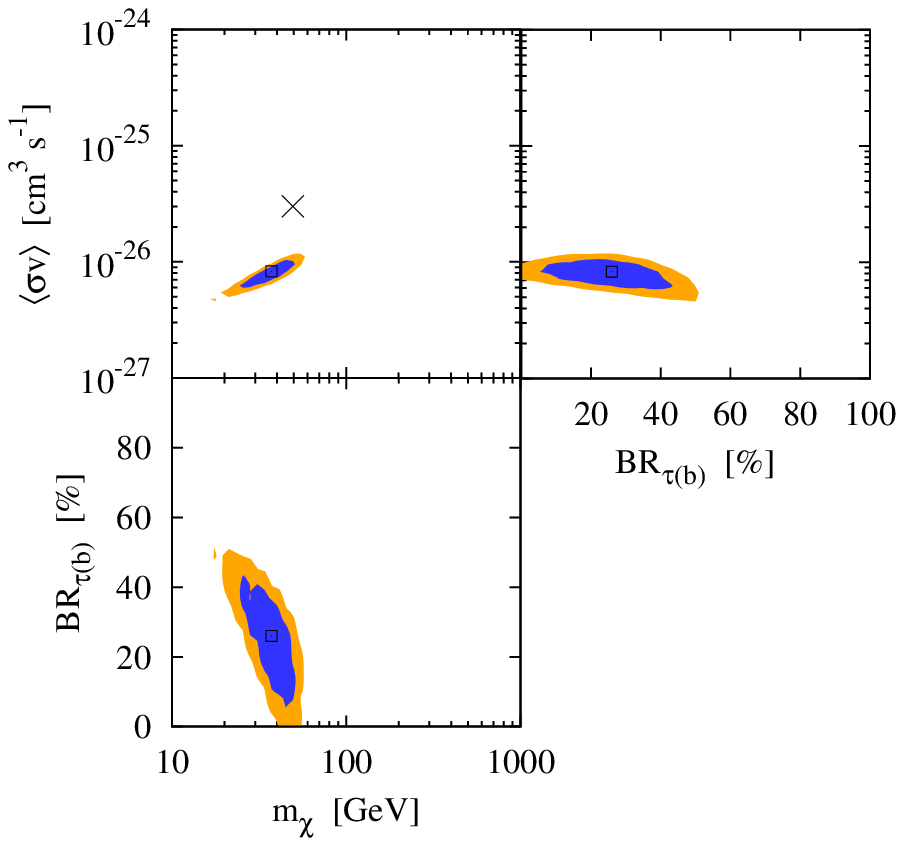}
\includegraphics[width=7.5cm]{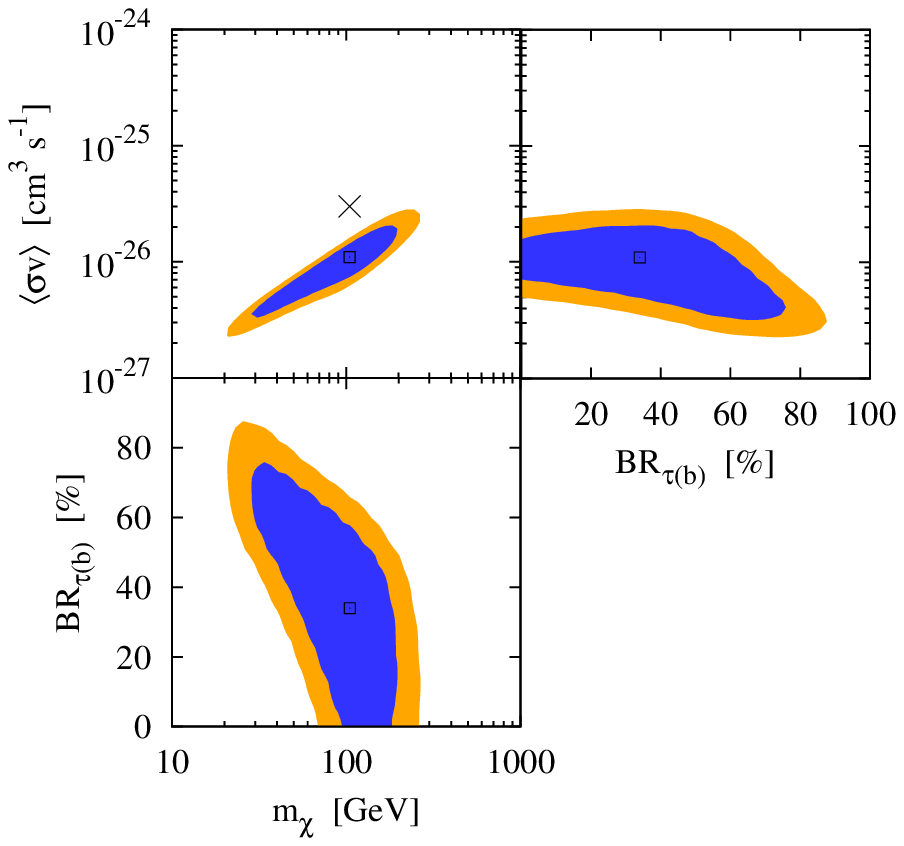}
\end{center}
\caption{{\it Fermi}--LAT abilities to constrain DM properties.  We
  assume the measured signal is due to DM annihilating into
  $\mu^+\mu^-$, but the fit is obtained assuming DM annihilates into
  either $\tau^+\tau^-$ or $b\bar b$.  We assume two DM masses:
  $m_\chi=50$~GeV (left panels) and $m_\chi=105$~GeV (right
  panels). Dark blue (light orange) regions represent the $68\%$~CL
  ($90\%$~CL) contours.  See the text for the rest of the
  parameters. The black cross in the left-top panel in each plot
  indicates the values of the parameters for the simulated observed
  ``data''.  Note that the other panels have no cross as they lie
  outside the parameter space of the simulated observed ``data''.  The
  squares indicate the best-fit point.}
\label{Fig:mubtau}
\end{figure}
In figure~\ref{Fig:mubtau} we assume that DM actually annihilates into
$\mu^+\mu^-$ pairs, but we analyze the data assuming DM annihilations
into either $\tau^+\tau^-$ or $b\bar b$,
for $\langle\sigma v\rangle=3\cdot 10^{-26}$~cm$^3$~s$^{-1}$
and for two DM masses: $m_\chi =
50$~GeV (left panels) and $m_\chi = 105$~GeV (right panels).
Again, the black crosses indicate the values of the parameters for the
simulated observed ``data''; the squares indicate the best-fit points.
Na\"{\i}vely, one would
expect that the $\mu^+\mu^-$ (leptonic) channel is identified as being
closer to the $\tau^+\tau^-$ (leptonic) channel than to the $b\bar b$
(hadronic) channel.
Let us remember that DM annihilation channels are
commonly classified into two broad classes: hadronic and leptonic
channels. Leptonic channels typically give rise to a harder
spectrum and, in particular for $e^+e^-$ and $\mu^+\mu^-$, the
cutoff is be very sharp and around a maximum energy (i.e. the mass of the
DM particle in the case on annihilating DM).
Contrary to what was expected, the reconstructed composition of the
annihilation channels tends to be dominated by $b\bar b$, instead of
$\tau^+\tau^-$. Indeed, the
contribution due to ICS in the case of the $\mu^+\mu^-$ (and also
the $e^+e^-$) channel could substantially alter the
different prompt spectra.
Hence, when taking into account the contribution of
ICS to the gamma-ray spectrum, the annihilation channels cannot be
generically classified as hadronic or leptonic, as DM annihilations
into $\mu^+\mu^-$ pairs are better reproduced with the $b\bar b$
channels than with the $\tau^+\tau^-$ channel.

\begin{figure}[ht!]
\begin{center}
\includegraphics[width=7.5cm]{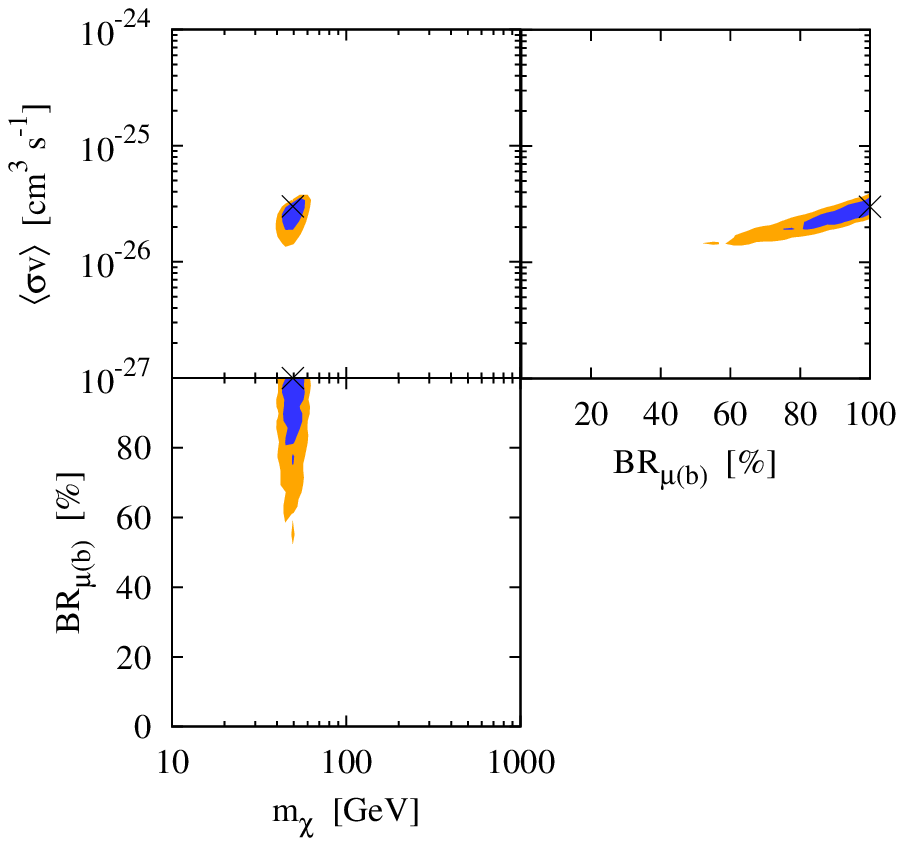}
\includegraphics[width=7.5cm]{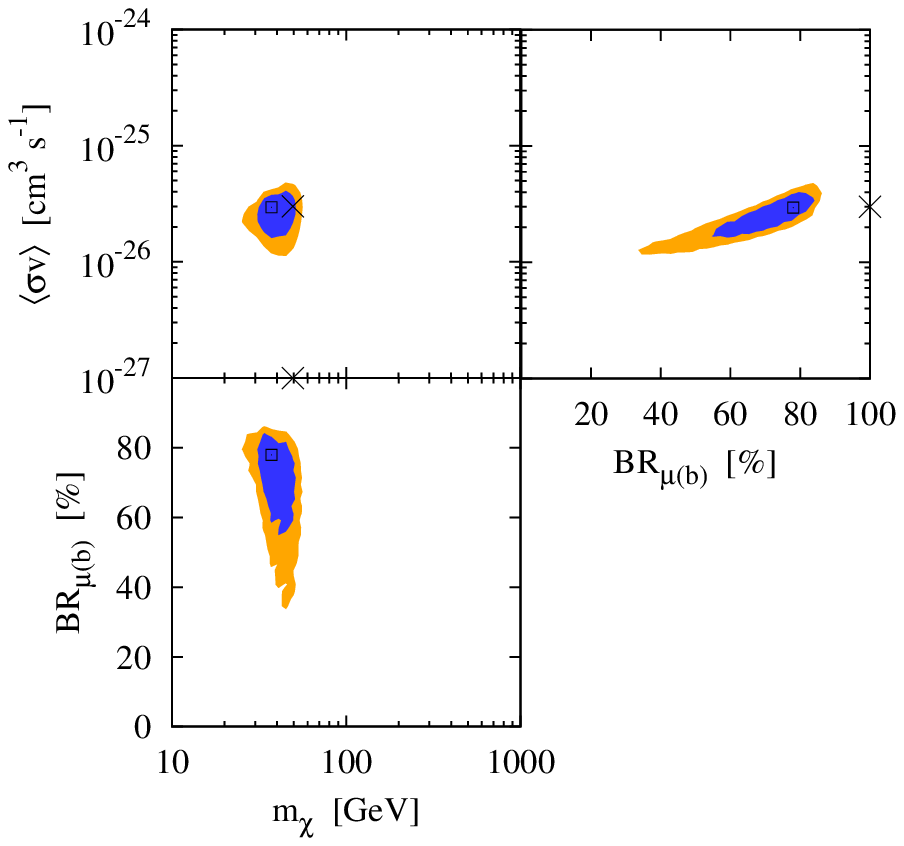}
\end{center}
\caption{{\it Fermi}--LAT abilities to constrain DM properties.  We
  assume the measured signal is due to DM annihilating into
  $\mu^+\mu^-$ and the fit is obtained assuming DM annihilates into
  $\mu^+\mu^-$ or $b\bar b$.  We assume ICS+prompt photons (left
  panels) or only prompt photons (right panels) for the reconstructed
  signal, for $m_\chi=50$~GeV.  Dark blue (light orange) regions
  represent the $68\%$~CL ($90\%$~CL) contours.  See
  the text for the rest of the parameters.  The black
  crosses indicate the values of the parameters for the simulated
  observed ``data''.  The squares in the right panels indicate the
  best-fit point.}
\label{Fig:mumub}
\end{figure}
The results just discussed can be illustrated in a different way by
analyzing the simulated observed signal ``data'' from DM annihilation
into $\mu^+\mu^-$ assuming DM annihilates into either $\mu^+\mu^-$
or $b\bar b$.  This is depicted in figure~\ref{Fig:mumub} where we show
the results for the case that we try to reconstruct the full signal
(prompt and ICS) generated by a $50$ GeV DM particle adding
the ICS contribution (left panels) or with only prompt photons (right
panels). As can be seen in the left panels, if ICS is taken into
account, DM properties can be reconstructed with good precision.
However, if the ICS contribution is not added to the simulated signal
events (the simulated observed ``data'' always has the ICS included),
DM annihilation into a pure $\mu^+\mu^-$ channel would be excluded at
more $90\%$~CL, providing thus a completely wrong result.

\section{Conclusions}
In this work we have studied the abilities of the {\it Fermi}--LAT
instrument to constrain Dark Matter
properties by using the current and future observations of gamma-rays
from the Galactic Center produced by DM annihilations.  Unlike
previous works, we also take into account the contribution to the
gamma-ray spectrum from ICS of electrons and positrons produced in DM
annihilations off the ambient photon background.
We show that the inclusion of the ICS contribution for
hadronic channels and for the $\tau^+\tau^-$ channel does not give
rise to important differences in the reconstruction process.
This is not
the case if DM annihilates into the $\mu^+\mu^-$ and the $e^+e^-$ channel.
In this latter case, adding the ICS
contribution to the prompt gamma-ray spectrum turns out to be crucial in order
not to obtain completely wrong results.

On the other hand, we found that for Dark Matter masses below $\sim 200$ GeV and for typical thermal annihilation cross sections,
it will be possible to obtain stringent bounds on the
Dark Matter properties such as its mass, annihilation cross section and annihilation channels.


\end{document}